Regular Article

Theoretical Physics

# Short-duration gamma-ray bursts from Kerr–Newman black hole mergers

Shad Ali[a]

Department of Astronomy, Xiamen University, Xiamen 361005, Fujian, China

**Abstract** Black hole (BH) mergers are natural sources of gravitational waves (GWs) and are possibly associated with electromagnetic events. Such events from a charged rotating BH with an accretion on to it could be more energetic and ultra-short-lived if the magnetic force dominates the accretion process because the attraction of ionized fluid with a strong magnetic field around the rotating BH further amplifies the acceleration of the charged particle via a gyromagnetic effect. Thus a stronger magnetic field and gravitational pull will provide an inward force to any fluid displaced in the radial direction and move it toward the axis of rotation with an increasing velocity. After many twists during rotation and the existence of restoring agents, Such events could produce a narrow intense jet starts in the form of Poynting flux along the axis of rotation resembling the Blandford–Znajek (BZ) mechanism. We investigated a charged rotating BH and obtained characteristic results (e.g., the remnant mass, magnetic field strength, luminosity, opening angle, viewing angle, and variation of viewing angle on the SGRB luminosity detection) that have a nice coincidence with rare events having GW associated with EM counterparts. This study gives a new insight into events with a strongly magnetized disk dominating the accretion process of energy extraction.

## 1 Introduction

Gamma-ray bursts (GRBs) are the most luminous and short-lived explosions e.g., [1–5]. This prompt emission is followed by a long-lived afterglow emission with energies ranging from X-rays to radio bands e.g., [6–11]. They can be classified as long-duration GRBs (LGRBs), $T_{90} > 2$ s; e.g., [12–18] and short-duration GRBs (SGRBs), $T_{90} < 2$ s; e.g. [19–21]. LGRBs are claimed to be emitted by the collapses to form a neutron star (NS) or a black hole (BH). The collapsar model is widely accepted as the standard model to describe the LGRB and ultra-LGRB emission e.g., [22–28], whereas, SGRBs are expected to originate from NS-NS or BH−NS mergers e.g., [29–36].

The double BHs mergers were not expected to power GRBs until a plausible SGRB associated with the binary BH (BBH) merger event GW150914 was first predicted by the Fermi-GBM [37] as a weak transient above 50 keV with a delay of 0.4s after the gravitational wave (GW) event with false alarm of 0.0022. This weak transient from the BBH merger lasted for 1s. In the source frame, the masses of the component BHs were observed as $m_1 = 36^{+5}_{-4}$ $M_\odot$ and $m_2 = 29^{+4}_{-4}$ $M_\odot$. After the merger, a remnant mass equal to $M = 62^{+4}_{-4}$ $M_\odot$ was observed whereas the mass $3.0^{+0.5}_{-0.5}$ $M_\odot$ is radiated as GWs associated with an unexpected electromagnetic event. The final BH spin parameter inferred from general relativity (GR) was $a = 0.67^{+0.05}_{-0.07}$. The source of this event is constrained to an annulus section of 610 deg$^2$, primarily in the southern hemisphere at the luminosity distance $410^{+160}_{-180}$ Mpc corresponding to red-shift $z = 0.09^{+0.03}_{-0.04}$ e.g., [38,39].

After predicting the event GW150914 as the progenitor of GW-GRB co-emission, many researchers put forward different models to explain the physics of this event e.g., see [40–47]. A hyperaccretion disc around a stellar-mass BH is considered one of the plausible central engines of GRBs e.g., for a review see [48]. The jets launched by the hyperaccretion disc are possibly powered by the neutrino annihilation process [23,49–55] or BZ mechanism [56]. It is worth noting that the neutrino annihilation process is effectively only for the central BH mass less than about 50 $M_\odot$ [57]. Janiuk et al. [46] studied the hyperaccretion disc. It is proposed that a collapsing massive star (making a BH) merges with another BH in a close binary leading to a weak transient due to the accretion onto the final BH with a moderate spin. The main mis-

[a] e-mail: shad.ali88@yahoo.com (corresponding author)

Springer

understandings in Janiuk's work are due to the time assumption, mass, and magnetic field of BBHs. First, Janiuk relayed the time assumption of $t = 2000M (\approx 0.6$ s$) \neq 1$ s. Second, their simulated data represents a mass of $62 M_\odot$ rather than a total mass of $65 M_\odot$. Third, the power and luminosity available through the Blandford–Znajek process at the horizon and in polar regions are considered negligible due to low spin and magnetization. Ignoring the magnetic field strength, they found the emitted power $\dot{E} = 1.1 \times 10^{50}$ erg s$^{-1}$ for $a = 0.8$ and $3.7 \times 10^{51}$ erg s$^{-1}$ for $a = 0.9$. The observed luminosity is much smaller as compared to these luminosities [37]. If the central BH rotates with a poloidal magnetic field threading to its horizon may drive a powerful relativistic jet by a process resembling the BZ mechanism [41,44]. Similarly, Khan [58] also performed magnetohydrodynamic simulations using an accretion disc onto the non-spinning BH binaries and investigated several properties of the event GW150914. We examined the event GW150914 by considering a BBH merger creating a KNBH as their final product with a magnetized disc emitting the rotational energy via narrow Poynting flux. We investigated their characteristic properties that could help the reader understand the physics of a jet possibly emitting from the merging binaries with a strong magnetized disk.

The structure of this paper is such that in Sect. 2, we present our model with a sketch of jet emission from a KNBH with a strong magnetic field around it that could be presented by Maxwell's EM equation. Such magnetic fields around a spinning BH could be the main source of GW associated with short-lived EM counterparts. In Sect. 3, we discuss the basic condition to set an instability in surrounding matter that leads to turbulence. In Sect. 4, we discussed the role of the gyromagnetic effect, and energy extraction from the final BH. This is the main part of adopting our model and the basic condition of instability in a suitable relation for the energy interpretation from the final BH. In Sect. 5, we considered the first observed event GW150914 in our model and obtained the luminosity associated with the emission of energy, the opening angle of the jet, and the effect of the viewing angle on the luminosity are discussed. Finally, some remarks and conclusions derived from this work are summarized in Sect. 6.

## 2 Model

Angular momentum and charge of BH retained by the magnetosphere are quantities creating a strong magnetic field around a spinning BH. The merger of such BHs could be naturally the progenitor of a new spinning BH with a stronger magnetic field that could extract the rotational energy from BH as an intense burst on a short time scale rather than by the accretion flow. In Refs. [31,59] have shown that a magnetic barrier will build up while the radial magnetic force counteracts the gravitational pull of the BH. Due to this reason, the rotational energy emission from the final BHs will have less chance to break the magnetic field along the radial direction. This energy of BH will act as a source of instabilities that tends to propagate along the axis of rotation due to a strong magnetic barrier and turbulent dynamo action while the gravitational pull along the radial direction. It could also provide further amplification of magnetic field( up to $\sim 10^{16}$ G) to excite electromagnetic waves in the form of a BZ jet e.g., [56,60]. The fact is that a charged spinning BH with a strong magnetic field and its dynamics could play an important role in accumulating magnetic flux e.g., [61].

The post-merger BH might have an accretion disk (formed during the merging process when material falls during the core collapse) that could be charged by friction. Such a spinning charged BH (Kerr–Newman) can in principle carry a force-free magnetosphere like magnetized spinning NSs (pulsars). they can sustain a global charge due to the spatial distribution of the charge density demanded to maintain a co-rotating force-free magnetosphere. If at least one of these merging BHs has some charge that is possibly retained by the magnetosphere of the BH, then it will strengthen the magnetic field around by induction as can be seen from the Maxwell equation

$$\nabla \times \boldsymbol{E} = -\partial_t \boldsymbol{B}, \quad (1)$$

where **E** is the electric field intensity of the BH charge and **B** is the magnetic induction intensity. Using Stokes's theorem, we can write this as

$$\oint_\Gamma \boldsymbol{E} ds = -\partial_t (Magnetic\ flux\ through\ S), \quad (2)$$

where $\Gamma$ is the closed curve bounding the surface $S$. The integral on the left is the electromotive force (EMF) and on the right is the rate of change of the magnetic flux, while the negative sign $(-)$ shows the direction of EMF and magnetic flux concerning each other.

Among several mechanisms, our model used Magneto-rotational instability (MRI), which relates significantly to turbulence in accretion disks. A hyper-accretion disk around a stellar-mass BH is a plausible source of the central engine to jet the GRBs as claimed by many authors but the main question is how to set the turbulence inside the hyper-accretion disk that results in jet GRBs. In search of the fact, the magnetic-field-driven disturbance called Magneto-Rotational Instability (MRI) is claimed to operate in hyper-accretion disks actively. It causes the angular momentum transport [61–63]. The MRI produces random motion (where the fluid moves faster toward the BH and slower outward) that generates fluctuations in both the magnetic field and the flow of the fluids. In such cases, the magnetic field lines restrict the outward displaced fluid and ensure the rigid rotation of the fluid that increases with time along the field direction to assume a cylindrical geometry. Due to these fluctuations and rigid rotations, each additional twist causes an increase in the cylindrical geometry. After many twists and restricted cross-sectional area, a narrow intense jet starts like a Poynting flux resembling the BZ mechanism as shown in Fig. 1.

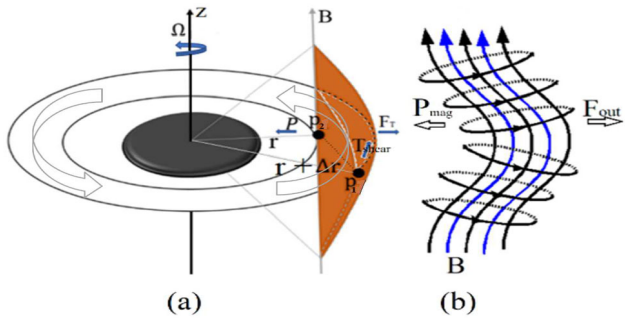

**Fig. 1** **a** Sketch for turbulence in disk due to an MRI-type instability. The fluid displaced from $p_1$ to $p_2$ feels a magnetic tension due to instability as restoring force while the shear component of the magnetic field spins up the BH so, the fluid will tend toward the axis of rotation. **b** The dynamics of the magnetic field lines under tension

By using Maxwell's stress tensor, the turbulence driven by the MRI can be described as [47]

$$T_{ij} = \epsilon_o \left( E_i E_j - \frac{\delta_{ij} E^2}{2} \right) + \frac{1}{\mu_o} \left( B_i B_j - \frac{\delta_{ij} B^2}{2} \right). \quad (3)$$

A BH can be completely described by its mass $M$, angular momentum $J$, and charge $q$. The first two quantities have been measured for stellar and supermassive BHs with various observations, but the charge is often neglected and implicitly set identically to zero. However, both classical and relativistic processes can lead to a small non-zero charge of black holes [64]. The charge of BH could be associated with the surrounding material making a disk. Due to the rotation property of the remnant BH, it is greatly expected that the charge contribution could be added to the magnetic fields so, to get a clearer insight into the turbulence, we decompose Maxwell's stress tensor into diagonal and off-diagonal components as

$$\delta_{ij} = \begin{cases} 1, & \text{if } i = j \\ 0, & \text{if } i \neq j. \end{cases} \quad (4)$$

So, for $i = j$, Maxwell's stress tensor becomes

$$P_{mag} = \frac{1}{2\mu_o} B^2, \quad (5)$$

where $P_{\text{mag}}$ is the restoring force on the perturbed fluid elements. For $i \neq j$, it gives the shear stress

$$T_{shear} = \frac{1}{\mu_o} B_{ij}. \quad (6)$$

In a differential rotating system, this tension force can lead the system to instability (Figs. 2, 3). The magnetic field

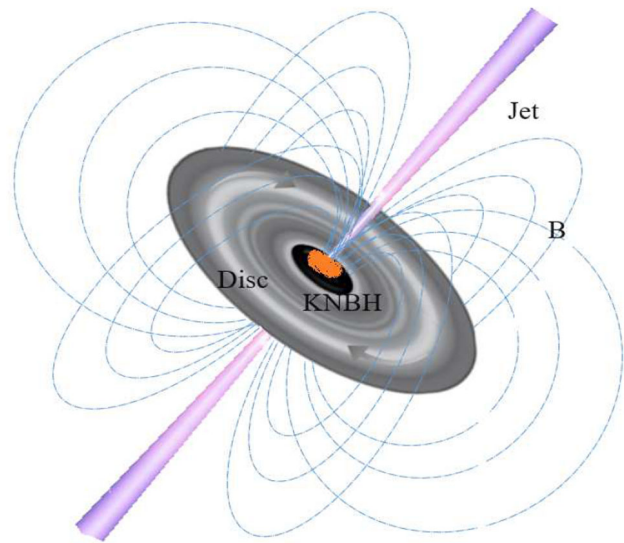

**Fig. 2** A sketch of KNBH with magnetized accretion disc emitting a narrow GRB by BZ jets

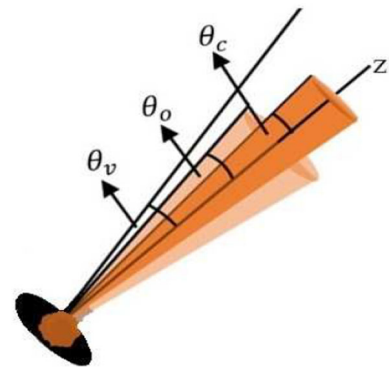

**Fig. 3** A close look at the jet along the axis of rotation from KNBH with magnetized accretion disc. The core angle $\theta_c$, opening angle $\theta_o$, and angle $\theta_v$ (the angle that the outgoing photon makes with the normal to the jet surface.) are shown from the axis of rotation [66–68]

is an effective source for this shear-type instability leading to turbulence in the accretion disc resulting in an inflow or outflow of fluid [65].

## 3 Basics condition for instability

Before demonstrating the remnant mass and the emitted energy from the merger system, we need some boundary conditions for instability caused by the merging process to overcome the system's stability. Various forces act upon the particles of the fluid in an accretion disc at different radii, so the fluid particles are moving at variable speeds [47]. As a result, the restoring force $P_{\text{mag}}$ along the gravitational forces on the fluid provides the potential required in a circular motion of the disc with an angular velocity $\Omega$ (or a calm Keplerian disc) and the shear force is responsible for seeding the instability at the instant of merging. This means that if a particle of the fluid is displaced from its position, then it will be acted upon by three forces i.e., gravitational and restoring magnetic tension toward the original position of the particle and the shear force along the diagonal but overall motion of the particle will be retained along the circular path. As long as the mass accretion increases, the instability increases, but due to rotational motion and magnetic restoring force, the displaced fluid tends toward the axis of rotation. Thus the crucial role of the magnetic field in shaping instability and fluctuation needs to be considered for understanding the basic process. Let the restoring magnetic tension acting on a particle is

$$P_{\text{mag}} = \frac{B^2}{4\pi\rho R} \approx \left(\frac{c_{\text{Al}}}{\lambda}\right)^2 \Delta r, \tag{7}$$

where $R = \lambda^2/\Delta r$ is the radius of magnetic field curvature with $\lambda$ as the perturbation wavelength in the vertical direction, $\rho$ is the density, and $c_{\text{Al}} \equiv \sqrt{B^2/4\pi\rho}$ as the speed of Alfvén wave. As the Alfvén propagates in the direction of the magnetic field. So, the motion of the ions/particle and the perturbation of the magnetic field are transverse to the direction of propagation [69]. This means that if the propagation of the magnetic field is in the radial direction the emitted energy (flux) will be perpendicular to it. From Eq. (7), the displacement $\Delta r$ at conserved angular velocity $\Omega = \sqrt{GM/r^3}$, $M$ is the total mass. The growth rate of centrifugal force for a unit mass is $\Delta F_c \sim \Omega^2 \Delta r$. It will result in the decrease of the gravitational force on unit mass by the amount

$$\Delta F_g \approx \frac{2GM}{r^3} \Delta r \approx 2\Omega^2 \Delta r. \tag{8}$$

Thus the total inward force acting on a single perturbed particle is $\sim 3\Omega^2 \Delta r$. For an instability to cause turbulence, the outward force must be dominated over the magnetic tension i.e., we can use the inequality between total force and Eq. (7) to represent the domination of magnetic tension,

$$3\Omega^2 > \left(\frac{c_{\text{Al}}}{\lambda}\right)^2 \approx \lambda > \frac{c_{\text{Al}}}{\Omega}, \tag{9}$$

Here, one can use angular velocity and Alfvén waves speed values to get a typical wavelength value that must satisfy the instability condition. In the case of rotating BH the angular velocity is defined as

$$\Omega = \sqrt{\frac{GM}{r^3}} \left(1 + \frac{a_*}{(8\hat{r}^3)^{\frac{1}{2}}}\right)^{-1} \tag{10}$$

here,

$$a_* = \frac{c^2 a}{GM} = \frac{cJ}{GM^2}, \qquad a = \frac{J}{Mc}$$

and $\hat{r} = \frac{r}{r_g}$. So, we can write as

$$\Omega = 2\sqrt{\frac{GM}{r^3}} \left(4 + \frac{\sqrt{2}cJ}{GM^2}\right)^{-1} \tag{11}$$

The condition of instability becomes

$$\lambda > \frac{\left(4 + \frac{\sqrt{2}cJ}{GM^2}\right) c_{Al}}{2\sqrt{\frac{GM}{r^3}}} \tag{12}$$

Let us consider the event $GW150914$ with $r = 10^6$cm and other typical values, we can estimate the instability condition as

$$\lambda = (1.1 \times 10^{-4} \text{sec}) c_{Al} \tag{13}$$

here the $c_{Al}$ depends on the mass density and the magnetic field strength. Near the magnetosphere, the Alfvén speed could rise to $10^8$cm/s with a wavelength up to $10^4$cm. Here, it is important to take $v_A \ll c$ because at $v_v \approx c$ the Alfvén wave becomes an ordinary electromagnetic wave. The growth rate of instability from the time scale is

$$\tau = \frac{\pi\lambda}{2c_{Al}} = \frac{\pi}{2\Omega} \approx 1.7 \times 10^{-4} \text{sec} > \frac{1}{\Omega}. \tag{14}$$

This equation shows that the growth rate of the instability depends on angular velocity and is independent of the magnetic field, i.e., the greater the angular velocity, the smaller the growth time of the turbulence, and vice versa. The magnetic field is an agent of seeding instability in the accreting fluid of the disc. At peak angular velocity, the instability grows in a very short time scale. Note that the instability sets in for perturbations with shorter wavelengths when the magnetic fields are weak.

## 4 Gyromagnetic effect and energy interpretations

The Schwarzschild and Reissner–Nordström radii of a charged BH are

$$r_s = \frac{2GM}{c^2} \text{ and } r_q = \frac{\sqrt{G}q}{c^2}, \tag{15}$$

where $M$ is the mass of BH. Comparing the above two equations, we can obtain the upper limit of the charge $q_c$ as

$$q_c = 2\sqrt{G}M = (6.1 \times 10^{31} \text{e.s.u}) \left(\frac{M}{10M_\odot}\right) \sim 2 \times 10^{22} \text{C}, \tag{16}$$

Such a limited quantity of BH's charge would significantly modify the magnetic field geometry to set the electric field that also contributes to the magnetic field via the gyromagnetic effect as given in Eq. (2). Due to BH charge contribution to the magnetic field during the gyromagnetic effect, we can follow the track of evaluating two quantities for describing a charged spinning BH as

$$\frac{dMc^2}{dt} = -P_L \quad \text{and} \quad \frac{dJ}{dt} = \tau, \quad (17)$$

where $P_L$ is the power due to angular momentum transportation, $\tau$ is the torque exerted by the magnetic field normal to the disc's surface, and the negative sign $(-)$ shows the power loss. In the hyperaccretion process, evolution should have the contribution from mass and angular momentum i.e., $\dot{M}E_{\text{ms}}$ and $\dot{M}J_{\text{ms}}$, where $\dot{M}$ is the accretion rate and $E_{\text{ms}}$ and $J_{\text{ms}}$ are the specific energy and angular momentum of the material accreted at the innermost radius, respectively. For KNBH, [41] investigated the magnetic dipole moment that drives Poynting flux during the merging process and claimed that the dissipation of Poynting flux at large radii would power SGRB if the BH has a charge of $10^{-5} - 10^{-4}$, and if the BH charge is $\sim 10^{-9} - 10^{-8}$ that will drive an FRB [44]. Let the disc of radius $r$ rotate with an angular velocity $\omega_F$, the change in torque and the rotational power of the disc is

$$\Delta\tau = -\frac{I}{2\pi}\Delta\psi \Rightarrow \Delta P_{\text{mag}} = -\omega_F \times \Delta\tau = \frac{\Delta\psi I \omega_F}{2\pi}, \quad (18)$$

here $I$ is the current due to the flow of charge and $\Delta\psi$ is the change in magnetic flux due to induction. If the loading region is unable to give rise to angular velocity due to large mass and inertia, then the power delivered by torque and the power of angular momentum transportation $\Delta P_L$ will be the same as $\Delta P_{\text{em}}$ from the disc, i.e.,

$$\Delta P_{em} = \Delta P_{\text{mag}} = \Delta P_{\text{L}} = \frac{\Delta\psi I \omega_F}{2\pi}. \quad (19)$$

By integrating, Eq. (19) could give the total BZ power as

$$P_{\text{mag}} = -\frac{1}{2\pi}\int \omega_F I d\psi \Rightarrow P_{\text{mag}} = -\frac{1}{2\pi}\int_h \omega_F I d\psi. \quad (20)$$

The subscript $h$ denotes the horizon. As $\omega$, $I$, and $\psi$ are constants along the magnetic field surface, it is easy to evaluate Eq. (20) on the horizon of the BH.

Consider the predicted merging event GW150914 resulting in a KNBH emitting some of its energy due to turbulence supported by the magnetic field of ionized fluid. So, combining the evolution relations of mass and angular momentum from Eq. (17) using Eq. (18), we get

$$\frac{dM}{dt} = \Omega_F \frac{dJ}{dt}, \quad (21)$$

here, the angular velocity of the field in Kerr Newman geometry can be written as [70]

$$\Omega_F = \frac{d\phi}{dt} = \frac{a}{r_+^2 + a^2}$$

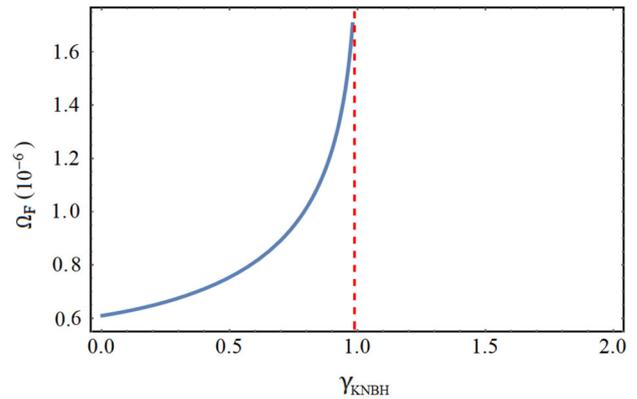

**Fig. 4** The plot of $\gamma_{\text{KNBH}}$ vs. $\Omega_F$ for a charged rotating BH with $M = 65M_\odot$ and $a = 0.67$

$$= \frac{J}{2M^3\left(1 - \frac{q^2}{2M^2} + \sqrt{1 - \frac{J^2}{M^4} - \frac{q^2}{M^2}}\right)}, \quad (22)$$

where $r_+ = M + \sqrt{M^2 - a^2 - q^2}$ is the KNBH horizon radius. KNBH has a non-zero angular momentum (spin) that imparts rotational energy to the BH, which can be tapped into. The magnetic field of such BH amplifies with the differential rotation. The charge-to-mass ratio $(q/M)$ in Eq. (22) is the gyromagnetic ratio of KNBH denoted by $\gamma_{\text{KNBH}}$ (the ratio of the magnetic moment to the angular momentum) of the Dirac electron.[1] Specifically, it is related to the radiation emitted by charged particles moving in the strong magnetic field around the BH.

For a KNBH, the magnetic dipole moment is $\mu = qa$, and angular momentum is $J = Ma$ so, $\gamma_{\text{KNBH}} = q/M > \gamma_{\text{classical}}$. The underlying fact of the gyromagnetic effects is that the nuclear and the electronic spins, as well as their orbital angular momenta, generate a magnetic moment parallel to the angular momentum of constant magnitude that can be transformed into a rotating coordinate system by Newton's second law of motion or Schrodinger's equation gives Larmor's theorem.[2]

In the context of magnetic fields, the gyromagnetic ratio determines the strength of the magnetic field produced by a spinning charged particle hence, it supports our statement that the BH charge will enhance the magnetic field of BH during its rotation. While solving Eqs. (21) and (22), one can't eliminate $M$ and $J$ to get a suitable solution. Hence, considering a fixed contribution of gyromagnetic ratio in Eq. (22)

---

[1] A standard electron that can be described by the Dirac theory also known as Larmor radiation that plays an important role in the emission of GRBs from a KNBH.

[2] The effect of a uniform magnetic field (produces periodic effects on a system of spins or particles) can be transformed away by going to a rotating coordinate system, provided the angular momentum of the system in the field direction is a constant of motion [71].

and neglecting the complex factors due to the smaller value of $\gamma_{KNBH}$ for the evaluation of KNBH can lead us to evaluate remnant mass by using the evaluation relation of $M$ and $J$. It is also a fact that the gyromagnetic ratio of KNBH and Dirac electron is always < 2 e.g., [72]. The relation of gyromagnetic effect vs. angular frequency ($\Omega_F$) is also drawn in Fig. 4. The plot shows an increase in the value of $\gamma_{KNBH}$ with a small increase of the $\Omega_F$ value, but as the $\gamma_{KNBH}$ crosses a certain limit, the $\Omega_F$ value increases without any significant increase in $\gamma_{KNBH}$ value. It shows a maximum effect of the gyromagnetic ratio on the $\Omega_F$ and the magnetic field strength. The gyromagnetic ratio approaches its maximum value i.e., $\sim 1$, which means the charge becomes neutral in a very short interval [44].

$$J^2 = M^4 \left(2x - x^2\right) + \gamma, \quad (23)$$

and by differentiating, we get

$$2J\frac{dJ}{dt} = 2M^4(1-x)\frac{dx}{dt} + (2x - x^2)M^3\frac{dM}{dt}, \quad (24)$$

hence, Eq. (21) gives

$$\frac{dM}{dt} = \frac{J}{2xM^3}\frac{dJ}{dt} \quad \Rightarrow \quad M \approx M_o\sqrt{\frac{x_o}{x}}, \quad (25)$$

here after the integration, we used the series expansion of the exponential term, to obtain Eq. (25). This relation could give the remnant mass of the merging binary. The difference may be due to the energy loss and the gyromagnetic effect during the merging process i.e. at the instant of BHs collision, the energy loss due to angular momentum transport lowers the spin of the remnant mass but the rotating ionized fluid may support the magnetic field via induction to maintain the stable spin. The subsequent electromagnetic radiation emission leads to spin-down of the BH to a stable spin that must be comparable to the particle's spin at ISCO. The amount of rotational energy extracted during this process is $M_o - M = \Delta M_o \approx 4.616\%$ is the expected mass energy extracted from a slowly rotating BHs [73]. This amount of energy is used to power the Poynting flux along the direction of the twisted magnetic field e.g., [56,74]. The remaining energy will be associated with the irreducible mass to increase the entropy of BH. The timescale for the GRB emission can be defined as the ratio of BH power to the power output. i.e., [38,74]

## 5 Jet luminosity and its variation

Bing Zhang [41] stated that if at least one BH in binary merger carries a certain amount of charge then the inspiral generates a current loop circuit inducing a dipole moment to power a Poynting flux. The orbital decay rate due to the GW emission is investigated as

$$\frac{da}{dt} = -\frac{2c}{5}\hat{a}^{-3}, \quad (26)$$

where $\hat{a}$ is the distance normalized to $2r_s$ and $c$ is the speed of light. The rapid evaluation of orbital separation (before merging) is considered to lead to a change in magnetic flux powering the Poynting flux. Next, by using the magnetic dipole moment, a relation between wind luminosity and GW luminosity is obtained as

$$L_w \approx 0.4 L_{GW} \hat{q}^2 \hat{a}^{-10}, \quad (27)$$

where $L_{GW} \approx 3.6 \times 10^{56} \hat{a}^{-5}$ erg s$^{-1}$ and $\hat{q}$ is the dimensionless parameter $\hat{q} \ll 1$. As for a charged rotating BH $\hat{a} < 1$, the luminosity value gets $\ll L_{GW}$ incompatible with the observed value [39]. At large radii, the dissipation energy of Poynting flux in the accretion flow would power SGRB if the BH has a charge of order $\hat{q} \sim 10^{-5} - 10^{-4}$. The main contradiction is that the electrodynamics description of the magnetosphere is left incomplete and, for simplification, the two BHs were supposed to have equal masses circulating in exactly circular inspiral. Furthermore, the magnetic field is taken as equivalent to that of a magnetar. Whereas in the case of the GW150914 event, the mass ratio of the BH is $\sim 0.80$. Bing Zhang modified this model in [75] after noticing that electric dipole radiations are more significant in charged Compact Binary Coalescence (cCBC) [76]. The luminosity for electric dipole radiation in general form for both masses was defined as

$$L_{e,dip} = \frac{c^5}{6G}(\hat{q}_1^2 + \hat{q}_2^2)\left(\frac{r_s(m_1)}{a}\right)^2\left(\frac{r_s(m_2)}{a}\right)^2 \quad (28)$$

here, $m_1$ and $m_2$ are the two masses of the binary and $a$ is the semi-major axis. The two points of this modified model are:

- The luminosity depends on the charges ($\hat{q}_1^2 + \hat{q}_2^2$) of the binary, the power increases whether the charges are the same or opposite.
- The Luminosity ratio to $L_{GW}$

$$\frac{L_{e,dip}}{L_{GW}} = \frac{5}{6}(\hat{q}_1^2 + \hat{q}_2^2)\left(\frac{[r_s(m_1)]^2[r_s(m_1)]^2}{[r_s(M_h)]^4}\right)\left(\frac{a}{r_s(M_h)}\right)$$

$$= \frac{5}{6}(\hat{q}_1^2 + \hat{q}_2^2)\left(\frac{M_r}{M}\right)^{\frac{2}{5}}\left(\frac{a}{r_s(M_h)}\right) \quad (29)$$

here, $M_h = M_r^{\frac{2}{5}} M^{\frac{3}{5}}$ is the horizon mass and $M_r$ as the reduced mass. The total separation energy is investigated by integrating the luminosity over time between the two separating points of the system

$$E_{e,dip} = \int_a^{a_{min}} \frac{L_{e,dip}}{\dot{a}} da$$
$$= \frac{5}{24} \ln\left(\frac{a}{a_{min}}\right)(\hat{q}_1^2 + \hat{q}_2^2) M_r c^2 \quad (30)$$

where the orbital decay rate due to GW loss is redefined as in Ref. [77]

$$\frac{da}{dt} = -\frac{64}{5} \frac{G^3 M_r M^2}{c^5 a^3} f(e) \quad (31)$$

with a correction factor $f(e) = \frac{1 + \frac{73}{24}e^2 + \frac{37}{96}e^4}{(1-e^2)^{\frac{7}{2}}} = 1$ for eccentricity $e = 0$. For comparison, the magnetic dipole moment energy is obtained as less than $L_{e.dip}$.

Similarly, [44] proposed a model showing that FRBs originate from the collapse magnetosphere of KNBH where the closed orbits of charged particles in the magnetosphere are unstable. The main focus of this model was the magnetosphere instability of a KNBH and the possible consequences that result in FRBs and their potential afterglow emission. In the present work, we focus on the predicted merger event GW150914, where the charge (in the form of a gyromagnetic effect) is crucial in amplifying the magnetic field. Using mass values, we calculated the magnetic field strength and, fixed our model to investigate the accretion rate and luminosity for this event as

$$L_{BZ} = f(a_*) \frac{B^2 r_s^2 c}{8\pi} \approx 9.2 \times 10^{50} \text{ erg s}^{-1}, \quad (32)$$

here, $a = 0.67$ is the BH's final spin the spin parameter $f(a_*) = 1 - \sqrt{\frac{1}{2}\left(1 + \sqrt{1-a^2}\right)} = 0.067$, and, $r_s \sim 10^6$ cm is used with other conventional units to obtain this luminosity value. We see that the obtained luminosity is greater than the observed luminosity. This may be due to several reasons. One can impose the beaming condition as [78]

$$L_{ob} = f_b L_{BZ} = 1.8 \times 10^{49} \text{erg s}^{-1}, \quad (33)$$

here, $f_b = 1 - \cos\theta_o = 0.0196$ with jet opening angle $\theta_o = 11.36^o$, which is consistent with the range of detectable GRBs opening angle. Compared to NS mergers, most BBH mergers may have no observable EM counterparts due to either there being no detected EM counterparts or having no transient at all. For several undetected EM counterparts, there may be several reasons like, some event may damp/ suppress the EM counterpart before its emission ( e.g., the absence of tidally disrupted material that is the most direct source of accretion to power a jet, availability of insufficient energy from the engine, absence of accretion disc, BH may be chargeless, feeding of materials from its accretion by BH) while some factors that make it undetected (including GW150914) even if it is emitted too (e.g., jet opening angle, orbital inclination (i.e., if the jet is not in the line of sight and if the jet is in the line of sight then the viewing angle of the observer has a great effect on the jet luminosity and hence its detection), or weakness of GBM system and its large localization uncertainty). In the case of GRBs from GW150914, the jet might not be emitted along the line of sight [79,80]. Moreover, for any hypothetical jet from BBH mergers, there is no ejecta to interact with (especially when there is a strong magnetic field); hence, the probability of observing EM radiation along the off-axis direction is much lower [66,80]. The gamma rays will likely be beamed during the GW-GRBs emission, while the GW emission is close to an isotropic propagation in all directions. Thus, it is difficult to rule out any particular model for a single event with both GW-EM emissions. Even if a model suggests a large flux, it is possible that the jet was beamed away from Earth, thus undetectable for GBM. This could be the case for many undetected events. Besides these, the existence of other transients and objects like asteroids during the event has also shown their effect on the EM event detection [81]. From the relativity, the observed luminosity at a given frequency is related to emitted luminosity as [82]

$$L_{ob} = 4\pi \delta^4 L_{em} \quad (34)$$

Here, the Doppler boosting factor (relativistic Doppler factor) is $\delta(\Gamma, \theta) = [\Gamma(1 - \beta \cos\theta_v)]^{-1}$ [83,84]. Compared to stationary sources, the Doppler factor modifies several measurable quantities for the observation. Using the values of $L_{em} = 9.2 \times 10^{50}$ erg s$^{-1}$, $\beta = \frac{v}{c} = 0.53$ [38], and $\Gamma$ range from $3.974 - 10.69$, we manipulate the relation between the observed luminosity and viewing angle as shown in Fig. 5.

The luminosity vs. viewing angle curve of Fig. 5 shows the apparent dynamics of the jet in the event GW150914. The intrinsic structure of the jet is the distribution of the angular energy within the polar angle. In contrast, the apparent structure of the jet is the observer's frame dependent and hence can be discussed by the Lorentz transformation. Using relativistic beaming one could explain the difference between these two distributions by using the Lorentz factor and the viewing angle $\theta_v$ [67]. In the above Fig. 5, $\theta_c = 0.055$ rad $\sim 3.15^o$ (point A) is the core angle for $\Gamma = 3.974$, $\beta \sim 0.53$ with the corresponding luminosity $L = 9.2 \times 10^{50}$ erg s$^{-1}$. From point A to point B, the curve shows the typically detectable luminosity with a viewing angle of 1.23 rad $\sim 70.5^o$ to the jet direction, $\Gamma = 6.93$, $\beta \sim 0.53$ with the corresponding luminosity $L = 1.017 \times 10^{50}$ erg s$^{-1}$. If the jet is emitted along the axis of rotation then at point C, we get the luminosity of $1.8 \times 10^{49}$ erg s$^{-1}$ with a viewing angle of 0.42 rad $\approx 24^o$ with $\Gamma = 10.69$. As $\Gamma > 1$, the observer receives emission from the whole emitting region (head of the jet) even if the $\theta_v$ is outside the confines of the intrinsic jet [68]. For the luminosity variation, the range of the Lorentz factor is 3.974–10.69, where the maximum value gives the core luminosity at a small viewing angle and the minimum value represents

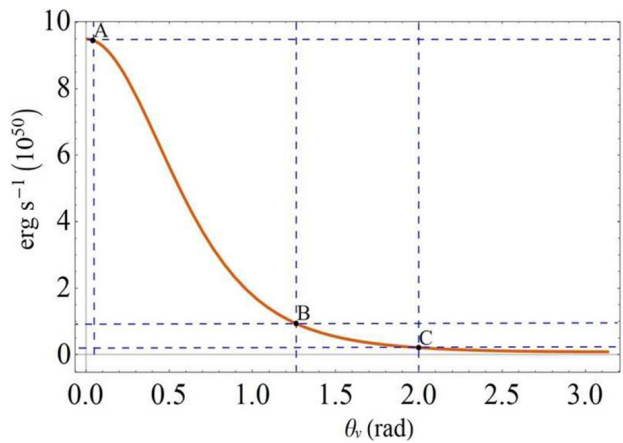

**Fig. 5** The plot of observed luminosity vs. viewing angle of the GRB jet from the event GW150914. Points A, B, and C show the luminosity at different viewing angles and the jet Lorentz factor

the observed luminosity of the jet at a luminosity distance of 410 MPc. here we can also see that $\theta_v \propto \frac{1}{\Gamma}$. Since the event GRB 170817A detection, the threshold distance is about 80 MPc with a threshold signal-to-noise ratio $\leq 8$. So, it seems that if the GW150914 event may have occurred but due to the lack of the GBM system the detection may not be observed. At a smaller angle, the luminosity is at its peak equivalent to core luminosity but as the $\theta_v$ value increases the luminosity decreases (represents a Gaussian curve). This shows that even if this event is detected, the observed luminosity $1.8 \times 10^{49}$ erg s$^{-1}$ is the minimum luminosity that could be detected from a source at a distance of 410 MPc. It is also claimed that the cosmological effects are small at such distances. Therefore, considering negligible cosmological redshift and apparent spectrum one can greatly understand the possible restrictions for many of the BBH events with no detectable EM counterparts.

## 6 Remarks and discussion

SGRBs are generally associated with the mergers of NS-NS or NS-BH binaries whereas, the BBH mergers were not expected to power GRBs until the possible weak transient from the event GW150914 (the first predicted weak transient might be associated with GWs from the BBH merger) [37]. If this event has happened from BBHs, then it will be a revolutionary trend in BH and GW physics. In this paper, we presented a possible scenario for the emission of co-emission of GW associated with the EM counterpart. This work could be a new insight into the new events that could not be explained by existing models.

We examine GW150914 as the BBH merger event where before merging the mutual attracting of a binary caused the surrounding matter to get inspiral around by making a hyper-accretion disc around the final BH. The matter gets ionized due to fraction and finally magnetized with rotation. Such magnetic fields are greatly affected by BH charge via the gyromagnetic effect. As a result, a strongly magnetized disc is generated. In such discs, the MRI-type instability causes turbulence with the magnetic fields serving as an agent to extract rotational energy. The existence of charge and its contribution to the magnetic flux is just the same as the total EMF in the system equal to the total change in magnetic flux through the disk's surface. From such a BH, the emission of energy along the radial direction has less chance due to the magnetic field barrier so, exactly after merging the extractible energy will be pushed toward the axis of rotation due to centripetal force and acting as a source of instability in the final BH. Next, further amplification of the magnetic field due to turbulent dynamo action($\sim 10^{16}$ G) excites the charged particle in the form of electromagnetic waves as a BZ jet e.g., [56,60]. The fact is that a charged spinning BH with a strong magnetic field and its dynamics could play an important role in accumulating magnetic flux e.g., [61]. So, the angular momentum and charge are two main factors for creating a strong magnetic field around spinning BHs that could extract rotational energy in the form of an intense ultra-short-lived burst.

Considering the instability of displacing the fluid, we obtained the magnetic pressure and shear force from the decomposition of Maxwell's tensor. These components clearly express the dynamics of the fluid in the accretion disc. Using these forces with gravitational force, we obtained the basic condition for MRI-type instability where the growth rate depends on angular velocity and is independent of the magnetic field strength. This means the magnetic field is an agent of seeding instability in the accreting fluid. Once the limitations to the stability are overcome, the instability sets in resulting in the emission of rotational energy. These limitations are just to overcome the potential and conserved quantities of motion for every single particle to maintain the stable circular motion of the fluid.

If the mass $M$, angular momentum $J$, and charge $q$ of a BH are satisfied with the basic condition of instability then one can demonstrate the energy and other quantities related to the extraction process of GRBs. The BH charge is crucial during the evaluation process in amplifying the magnetic field via the gyromagnetic effect. An analytic result from the $\Omega_F - \gamma_{KNBH}$ relation is shown in Fig. 4. This figure shows an increase in the $\gamma_{KNBH}$ value with a small increase in the $\Omega_F$ value but, as the $\gamma_{KNBH}$ crosses a certain limit, the $\Omega_F$ value increases without any significant increase in $\gamma_{KNBH}$ value that signifies the maximum effect of the gyromagnetic effect on $\Omega_F$ and the magnetic field strength. The maximum value of the gyromagnetic ratio is $\approx 1$, which means the charge becomes neutral as $\Omega_F$ approaches its peak value (that shows a very short interval of time). By considering the charge effect other than the gyromagnetic effect, one can't get the exact solution for Eq. (21) due to a problem in the elimination of mass $M$ and angular momentum $J$ of a KNBH. So, considering a fixed contribution of the gyromagnetic effect in the evaluation of KNBH could lead us to calculate the remnant mass from the assessment of $M$ and $J$.

Considering the event GW150914 as the merging of BBHs that possibly emits a portion of its rotational energy in the form of GW-EM energy. We evaluate the mass and angular momentum of the final BH with a gyromagnetic effect and investigate remnant mass equal to $M = 62^{+4}_{-4}\ M_\odot$ where the initial spin exactly after merging is found to be 0.736 and the final spin at the time of predicted GW-EM radiation emission was 0.67. Initially, the BBHs spun fast but at the instant of BBH's collision, the energy loss due to angular momentum transport lowers the spin of the remnant mass. Still, the rotating ionized fluid supports the magnetic field via induction to maintain the spin of the irreducible remnant mass. The energy extraction leads to a spin-down of the BH that is comparable to the particle's spin at ISCO. The amount of rotational energy extracted during this process is predicted to be $\sim 3\ M_\odot$ in the form of GWs associated with a weak transient. We investigated the magnetic field of order $\sim 10^{15}$ G to create turbulence. This is enough magnetic field to power the Poynting flux along the axis of rotation. In all this situation, we can't neglect the effect of BH's charge with the existence of a strong magnetic field that is a key factor in collimating the jet and extracting the rotational energy from BH. The maximum luminosity emitted during this process is found $9.2 \times 10^{50}$ erg s$^{-1}$ greater than the observed luminosity $1.8 \times 10^{49}$ erg s$^{-1}$ shows that if the jet is emitted along the axis of rotation but not along the line of sight. For this purpose, we analyzed the luminosity variation with the viewing angle shown in Fig. 5 and other effective factors. The luminosity of the emitted jet decreases with the increase of viewing angle and Lorentz factor. The viewing angle for the observed luminosity is found to be $24^o$. Examining the geometry and using the definition of viewing angle [85], one can predict the inclination angle of $166^o$. Most of the BBH mergers have no EM counterparts due to several restrictions but for a jet along the axis of rotation with a strong magnetic field, it must have no ejecta with which the emitted jet could interact so, there is a high probability of detection if the emitted jet is along the line of sight and as the jet goes away from the line of sight, its luminosity decreases and hence probability of its detection.

**Acknowledgements** I sincerely thank and appreciate the editor and anonymous referee's effort for a careful reading and helpful suggestions on this manuscript. These suggestions significantly improved the work done in this paper. I am also thankful to Prof. Tong Liu and Prof. Li Xue for their valuable discussion. This work is a part of my Postdoc study at Xiamen University and is supported by the China Postdoctoral Science Foundation.

**Funding** This work is supported by China postdoctoral Science foundation and Department of Astronomy of Xiamen University. There is no any other funding information.

**Data Availability Statement** Data will be made available on reasonable request. [Author's comment: The data underlying this article will be shared on reasonable request to the corresponding author.]

**Code Availability Statement** The manuscript has no associated code/software. [Author's comment: Code/Software sharing not applicable to this article as no code/software was generated.]